\documentclass{elsart}
\usepackage{graphicx}
\usepackage[usenames]{color}
\usepackage{amssymb}
\usepackage{amsmath}

\begin{document}
\newcommand{\aff}[2]{Dipartimento di Fisica dell'Universit\`a #1 e Sezione INFN, #2, Italy.}
\newcommand{\affuni}[2]{Dipartimento di Fisica dell'Universit\`a #1, #2, Italy.}
\newcommand{\affinfn}[2]{INFN Sezione di #1, #2, Italy.}
\newcommand{\affinfnm}[2]{INFN Sezione di #2, #2, Italy.}
\newcommand{\affinfnn}[2]{INFN Sezione di #1, #2, Italy.}
\newcommand{\affd}[1]{Dipartimento di Fisica dell'Universit\`a e Sezione INFN, #1, Italy.}

\newcommand{\dafne}	{DA$\Phi$NE }

\newcommand{\phirpi}	{\phi \rightarrow \rho \pi}
\newcommand{\phikk}	{\phi \rightarrow K \bar{K}}
\newcommand{\phippp}	{\phi \rightarrow \pi^+ \pi^- \pi^0}
\newcommand{\phietag}	{\phi \rightarrow \eta \gamma}

\newcommand{\etappeeg}	{\eta \rightarrow \pi^+ \pi^- e^+ e^- (\gamma)}
\newcommand{\etappee}	{\eta \rightarrow \pi^+ \pi^- e^+ e^- }
\newcommand{\pp}	{\pi^+ \pi^- }
\newcommand{\ee}	{e^+ e^- }
\newcommand{\ppee}	{\pi^+ \pi^- e^+ e^- }
\newcommand{\eeee}	{e^+ e^- e^+ e^- }

\newcommand{\etapp}	{\eta \rightarrow \pi^+ \pi^-}
\newcommand{\etappg}	{\eta \rightarrow \pi^+ \pi^- \gamma}
\newcommand{\etappp}	{\eta \rightarrow \pi^+ \pi^- \pi^0}
\newcommand{\etapOpOpO}	{\eta \rightarrow \pi^0 \pi^0 \pi^0}

\newcommand{\bharad}	{e^+ e^- \to e^+ e^- (\gamma)}
\newcommand{\eephietag}	{e^+ e^- \to \phi \to \eta \gamma}

\newcommand{\aphi}	{\mathcal{A}_{\phi}}
\newcommand{\spcp}	{\sin \phi \cos \phi}
\newcommand{\spcpp}	{\sin \phi \cos \phi>0}
\newcommand{\spcpm}	{\sin \phi \cos \phi<0}
\newcommand{\ltb}       {\langle \cos \theta_b \rangle}
\newcommand{\ltf}       {\langle \cos \theta_f \rangle}

\begin{frontmatter}

\title{Measurement of the branching ratio and search for a CP violating
       asymmetry in the \mathversion{bold}$\etappeeg$\mathversion{normal}
       decay at KLOE}

\collab{The KLOE Collaboration}

\author[Na,infnNa]{F.~Ambrosino},
\author[Frascati]{A.~Antonelli},
\author[Frascati]{M.~Antonelli},
\author[Roma2,infnRoma2]{F.~Archilli},
\author[Karlsruhe]{P.~Beltrame},
\author[Frascati]{G.~Bencivenni},
\author[Frascati]{S.~Bertolucci},
\author[Roma1,infnRoma1]{C.~Bini},
\author[Frascati]{C.~Bloise},
\author[Roma3,infnRoma3]{S.~Bocchetta},
\author[Frascati]{F.~Bossi},
\author[infnRoma3]{P.~Branchini},
\author[Frascati]{G.~Capon},
\author[Frascati]{T.~Capussela},
\author[Roma3,infnRoma3]{F.~Ceradini},
\author[Frascati]{P.~Ciambrone},
\author[Roma1]{F.~Crucianelli},
\author[Frascati]{E.~De~Lucia},
\author[Roma1,infnRoma1]{A.~De~Santis},
\author[Frascati]{P.~De~Simone},
\author[Roma1,infnRoma1]{G.~De~Zorzi},
\author[Karlsruhe]{A.~Denig},
\author[Roma1,infnRoma1]{A.~Di~Domenico},
\author[infnNa]{C.~Di~Donato},
\author[Roma3,infnRoma3]{B.~Di~Micco},
\author[Frascati]{M.~Dreucci},
\author[Frascati]{G.~Felici},
\author[Roma1,infnRoma1]{S.~Fiore},
\author[Roma1,infnRoma1]{P.~Franzini},
\author[Frascati]{C.~Gatti},
\author[Roma1,infnRoma1]{P.~Gauzzi},
\author[Frascati]{S.~Giovannella\corauthref{cor}},
\ead{simona.giovannella@lnf.infn.it}
\author[infnRoma3]{E.~Graziani},
\author[Frascati]{G.~Lanfranchi},
\author[Frascati,StonyBrook]{J.~Lee-Franzini},
\author[Karlsruhe]{D.~Leone},
\author[Frascati,Energ]{M.~Martini},
\author[Na,infnNa]{P.~Massarotti},
\author[Na,infnNa]{S.~Meola},
\author[Frascati]{S.~Miscetti},
\author[Frascati]{M.~Moulson},
\author[Frascati]{S.~M\"uller},
\author[Frascati]{F.~Murtas},
\author[Na,infnNa]{M.~Napolitano},
\author[Roma3,infnRoma3]{F.~Nguyen},
\author[Frascati]{M.~Palutan},
\author[infnRoma1]{E.~Pasqualucci},
\author[infnRoma3]{A.~Passeri},
\author[Frascati,Energ]{V.~Patera},
\author[Na,infnNa]{F.~Perfetto},
\author[Frascati]{P.~Santangelo},
\author[Frascati]{B.~Sciascia},
\author[Frascati]{T.~Spadaro},
\author[Roma1,infnRoma1]{M.~Testa},
\author[infnRoma3]{L.~Tortora},
\author[infnRoma1]{P.~Valente},
\author[Frascati]{G.~Venanzoni},
\author[Frascati,Energ]{R.~Versaci\corauthref{cor}},
\ead{roberto.versaci@lnf.infn.it}
\corauth[cor]{Corresponding author.}
\author[Frascati,Beijing]{G.~Xu}

\address[Frascati]{Laboratori Nazionali di Frascati dell'INFN, Frascati, Italy.}
\address[Karlsruhe]{Institut f\"ur Experimentelle Kernphysik, 
                    Universit\"at Karlsruhe, Germany.}
\address[Na]{Dipartimento di Scienze Fisiche dell'Universit\`a 
             ``Federico II'', Napoli, Italy}
\address[infnNa]{INFN Sezione di Napoli, Napoli, Italy}
\address[Energ]{Dipartimento di Energetica dell'Universit\`a 
                ``La Sapienza'', Roma, Italy.}
\address[Roma1]{\affuni{``La Sapienza''}{Roma}}
\address[infnRoma1]{\affinfnm{``La Sapienza''}{Roma}}
\address[Roma2]{\affuni{``Tor Vergata''}{Roma}}
\address[infnRoma2]{\affinfnn{Roma Tor Vergata}{Roma}}
\address[Roma3]{\affuni{``Roma Tre''}{Roma}}
\address[infnRoma3]{\affinfnn{Roma Tre}{Roma}}
\address[StonyBrook]{Physics Department, State University of New York 
                     at Stony Brook, USA.}
\address[Beijing]{Institute of High Energy Physics of Academica Sinica, 
                  Beijing, China.}

\begin{abstract}
We have studied the $\etappeeg$ decay using about 1.7 $fb^{-1}$ collected
by the KLOE experiment at the \dafne $\phi$-factory.
This corresponds to about 72 millions $\eta$ mesons produced in $\phi$
radiative decays.
We have measured the branching ratio, inclusive of radiative effects,
with 4\% accuracy:
$BR(\etappeeg) = (26.8 \pm 0.9_{Stat.}\pm 0.7_{Syst.}) \times 10^{-5}$.
We have obtained the first measurement of the CP-odd $\pi\pi-ee$ decay
planes angular asymmetry, 
$\aphi = (-0.6 \pm 2.5_{\ Stat.} \pm 1.8_{\ Syst.}) \times 10^{-2}$.
\end{abstract}

\end{frontmatter}

\section{Introduction}
\label{sec:introduction}
The decay of light pseudoscalar mesons, $\pi^0$, $\eta$ and $\eta'$,
proceeds via electromagnetic interaction and the radiative decays of
$\eta$ and $\eta'$ to pions allow to probe their electromagnetic
structure~\cite{Landsberg}. 
Conversion decays offer the possibility to measure precisely the virtual
photon 4-momentum via the invariant mass of the $\ee$ pair. 
The branching ratio for internal conversion decay of the $\eta$ meson,
$\etappee$, has been computed with different approaches, but until recently
both the theoretical and the experimental results were affected by large
uncertainties.
\\
The first calculation, based on pure QED, is 40 years old: 
$BR \sim 3 \times 10^{-4}$~\cite{Jarlskog}. 
The addition of $\pi\pi$ interaction treated with the Vector Dominance
Model gives $BR = 3.6 \times 10^{-4}$ with an error of about
10\%~\cite{Faessler}, while an approach based on chiral perturbation theory
that includes vector mesons 
gives $(3.2 \pm 0.3) \times 10^{-4}$~\cite{Picciotto}.
Recently, an approach based on the chiral effective Lagrangian
including $\pi\pi$ interactions has obtained a more precise result:
$BR = (2.99^{+ 0.06}_{-0.09}) \times 10^{-4}$~\cite{Borasoy}. 
\\
The $\etappee$ decay has been first observed by the CMD-2
experiment~\cite{Akhmetshin:2000bw}, giving 
$BR = (3.7^{+2.5}_{-1.8} \pm 3.0) \times 10^{-4}$, and has afterwards been
confirmed by the CELSIUS-WASA
experiment~\cite{Bargholtz:2006gz,Berlowski:2008zz}:
$BR = (4.3^{+2.0}_{-1.6} \pm 0.4) \times 10^{-4}$. 
The precision of these results does not allow to test different
models. 
\\
Recently, a possible CP violating mechanism (CPV), not directly related to
the most widely studied flavor changing neutral processes, has been
proposed. 
This mechanism could induce interference between electric and magnetic
decay amplitudes.
Such CPV effect could be tested in the decays of the pseudoscalar mesons
by measuring the polarization of the virtual photon and would result in an
asymmetry in the angle $\phi$ between the planes containing the $\ee$ and
the $\pp$ pairs in the meson rest frame, defined as:
$
\aphi =( \int_0^{\pi/2}    \frac{d\Gamma}{d\phi}d\phi -
         \int_{\pi/2}^{\pi} \frac{d\Gamma}{d\phi}d\phi ) /
       ( \int_0^{\pi/2}    \frac{d\Gamma}{d\phi}d\phi +
         \int_{\pi/2}^{\pi} \frac{d\Gamma}{d\phi}d\phi ) 
$.
This kind of asymmetry has been already studied
\cite{Sehgal:1992wm,Heiliger:1993qt,Elwood:1995xv,Elwood:1995dj,Ecker:2000nj} 
and observed \cite{AlaviHarati:1999ff,Lai:2000xf} in the decay of the $K_L$
meson.
In the $\eta$ decay this asymmetry is constrained, by
experimental~\cite{Ambrosino:2004ww} and Standard
Model~\cite{Jarlskog:1995ww} upper limits on the CP-violating decay 
$\etapp$, to be, at most of $\mathcal{O}(10^{-4})$ and
$\mathcal{O}(10^{-15})$, respectively. 
However, as pointed out in \cite{Geng:2002ua,Gao:2002gq}, it is possible in
case of  sources of CPV beyond the Standard Model which do not contribute
directly either to $\epsilon_K$ or to the neutron electric dipole moment,
$d_n$. 
In this case, $\aphi$ is predicted to be up to $\mathcal{O}(10^{-2})$, a
value reachable with the present statistics of $\eta$ mesons.

\section{The KLOE detector}
\label{sec:detector}
The KLOE experiment operates at \dafne, the Frascati $\phi$-factory. 
DA$\Phi$NE is an $e^+e^-$ collider running at a center of mass energy 
of $\sim 1020$~MeV, the mass of the $\phi$ meson. 
Equal energy positron and electron beams collide at an angle 
of $\pi$-25 mrad, producing $\phi$ mesons nearly at rest.
\\
The detector consists of a large cylindrical Drift Chamber, surrounded by a
lead-scintillating fiber ElectroMagnetic Calorimeter (EMC). 
A superconducting coil around the EMC provides a 0.52~T field.
The drift chamber~\cite{DCH}, 4~m in diameter and 3.3~m long, has 12,582
all-stereo tungsten sense wires and 37,746 aluminum field wires. 
The chamber shell is made of carbon fiber-epoxy composite and the gas used
is a 90\% helium, 10\% isobutane mixture. 
The position resolutions are $\sigma_{xy} \sim 150\ \mu$m and 
$\sigma_z \sim$~2 mm.
The momentum resolution is $\sigma(p_{\perp})/p_{\perp}\approx 0.4\%$.
Vertices are reconstructed with a spatial resolution of $\sim$ 3~mm.
The calorimeter~\cite{EMC} is divided into a barrel and two endcaps, for a
total of 88 modules, and covers 98\% of the solid angle. 
The modules are read out at both ends by photo-multipliers, both in
amplitude and time. 
The readout granularity is $\sim$\,(4.4 $\times$ 4.4)~cm$^2$, for a total
of 2440 cells arranged in five layers. 
The energy deposits are obtained from the signal amplitude while the
arrival times and the particles positions are obtained from the time
differences. 
Cells close in time and space are grouped into calorimeter clusters. 
The cluster energy $E$ is the sum of the cell energies.
The cluster time $T$ and position $\vec{R}$ are energy-weighed averages. 
Energy and time resolutions are $\sigma_E/E = 5.7\%/\sqrt{E\ {\rm(GeV)}}$ 
and  $\sigma_t = 57\ {\rm ps}/\sqrt{E\ {\rm(GeV)}} \oplus100\ {\rm ps}$, 
respectively.
The trigger \cite{TRG} uses both calorimeter and chamber information.
In this analysis the events are selected by the calorimeter trigger,
requiring two energy deposits with $E>50$ MeV for the barrel and $E>150$
MeV for the endcaps. 
A cosmic veto rejects events with at least two energy deposits above 30 MeV
in the outermost calorimeter layer. 
Data are then analyzed by an event classification filter \cite{NIMOffline},
which streams various categories of events in different output files.

\section{Event selection}
\label{sec:eventselection}
This analysis has been performed using 
 1,733 $pb^{-1}$ from the 2004-2005 dataset, 
   242 $pb^{-1}$ from the 2006 off-peak ($\sqrt{s}=1000\ MeV$) data,   
 3,447 $pb^{-1}$ of Monte Carlo (MC) simulating all $\phi$ decays,
50,506 $pb^{-1}$ of signal Monte Carlo.
Signal MC has been generated according to the matrix element in
\cite{Gao:2002gq}, with $BR = 4 \times 10^{-4}$ and having $\aphi = 0$.
All MC productions account for run by run variation of the main data-taking
parameters such as background conditions, detector response and beams
configuration.
Data-MC corrections for calorimeter clusters and tracking efficiency,
evaluated with radiative Bhabha events and $\phirpi$ samples respectively,
have been applied.
Effects of Final State Radiation (FSR) have been taken into account using
the PHOTOS MC package \cite{Barberio:1993qi,Golonka:2005pn}.
This package simulates the emission of FSR photons by any of the decay
products taking also into account possible interference between different
diagrams.
We have inserted PHOTOS in our Monte Carlo at the event generation level,
simulating FSR for the electrons, therefore our simulation fully accounts
for radiative decays.
\\
At KLOE, $\eta$ mesons are produced together with a monochromatic recoil
photon through the radiative decay $\phietag$ ($E_{\gamma} = 363\ MeV$).
In the considered data sample about $72 \times 10^{6}\ \eta$'s are present.
As first step of the analysis, a preselection is performed
requiring at least four tracks (two positive and two negative) coming
from the Interaction Point.
The Fiducial Volume is defined by a cylinder having radius $R = 4$ cm
and height $h = 20$ cm.
For each charge, the two tracks with the highest momenta are selected.
After track selection, one and only one neutral cluster, having energy
$E_{cl} \ge 250$ MeV and polar angle in the range $(23^\circ -157^\circ)$, 
is required. 
A cluster is defined neutral if it does not have any associated track
and has a time compatible with the photon time of flight.
After preselection, the signal is about 1.4\% of the sample.
\\
Mass assignment for each track is performed by either identifying a pion
decay from a kink in the track, or using the Time Of Flight (TOF) of the
particles.
For each track associated to a calorimeter cluster, the quantity 
$\Delta t = t_{track} - t_{cluster}$ in both electron ($\Delta t_e$) and
pion ($\Delta t_{\pi}$) hypothesis is evaluated; 
$t_{track}$ is defined as the length of the track divided by $\beta(m)c$.
The mass hypothesis minimizing $\Delta t$ is then chosen.
If two tracks of the same charge satisfy the same mass hypothesis, the
minimum $\Delta t_e$ identifies the electron.
When only one of the same charge tracks is identified as pion (electron), the
other one is assumed to be electron (pion).
For the remaining events the tracks with higher momentum are assumed to be
pion.
\\
To improve the energy and momentum resolution, a kinematic fit is performed
imposing the four-momentum conservation and the TOF of the photon.
A very loose cut on the $\chi^2$ of the kinematic fit is applied in order
to discard poorly reconstructed events.

\section{Background rejection}
\label{sec:backgroundrejection}
\begin{figure}
  \begin{center}
    \includegraphics*[width=7cm]{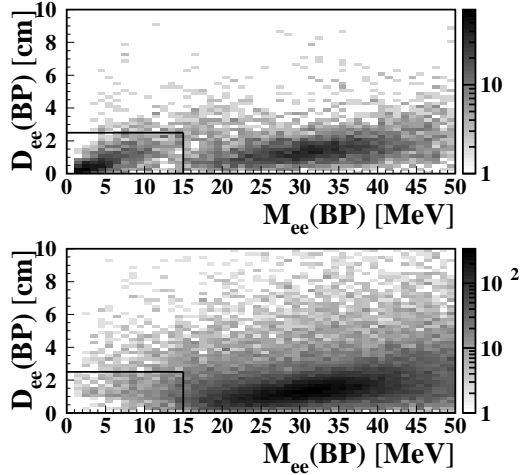}
  \end{center}
  \caption{$D_{ee}$ vs $M_{ee}$ evaluated at the beam pipe for $\phi$
           background (top panel) and MC signal (bottom panel). 
           Events in the box 
           $M_{ee}(BP) < 15\ MeV\ \cap\ D_{ee}(BP) < 2.5\ cm$ are rejected.}
  \label{fig:deemee}
\end{figure}

\begin{figure}
  \begin{center}
    \includegraphics*[width=7cm]{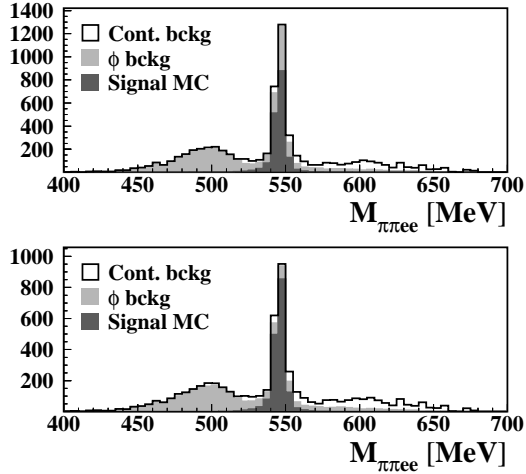}
  \end{center}
  \caption{Spectrum of the $\ppee$ invariant mass after the cuts on the
           momenta (top panel) and after the cut to reject events with
           photon conversions (bottom panel) have been applied.
           The black histogram is the expected distribution, i.e. signal MC
           (dark grey), $\phi$ background (light grey) and continuum
           background (white).
           The three samples have been normalized according to their
           luminosity.} 
  \label{fig:mppee1}
\end{figure}

Two sources of background have to be distinguished:
\begin{enumerate}
\item{$\phi$ background:
  \\
  this is mainly due to $\phippp$ events (with $\pi^0$ Dalitz decay) and to 
  $\phietag$ events either with $\etappp$ (with $\pi^0$ Dalitz decay) or
  with $\etappg$ (with photon conversion on the beam pipe).
  Note that this last background has the same signature as the signal.
  Background from $\phikk$ is also present at the preselection level.
  \\
}
\item{Continuum background:
  \\
  this is due to $\bharad$ events with photon conversions,
  split tracks or interactions with some material in the region of
  \dafne quadrupoles inside KLOE.
  Because of poor MC statistics, this background has been studied using
  off-peak data taken at $\sqrt{s}=1\ GeV$, where $\phi$ decays are
  negligible. 
  This sample has been properly normalized for luminosity and $\sqrt{s}$
  behaviour.
}
\end{enumerate}
A first background rejection is performed cutting on the sum of the
absolute value of the momenta of the two particles having the highest
momenta and opposite charge,  
$s2p = |\vec{p}(\mbox{p}_{Max}^{+})|~+~|\vec{p}(\mbox{p}_{Max}^{-})|$, 
and cutting on the sum of the absolute value of the momenta of the four
selected tracks, $s4p = \sum_1^4 |\vec{p}_i|$. 
It is required that:
$(270 < s2p < 460)\ MeV$ and $(450 < s4p < 600)\ MeV$.
\\
To reject the events due to photon conversion on the beam pipe, we
extrapolate backward the $\ee$ candidate tracks, down to the intersection
with the BP, and compute there the invariant mass ($M_{ee}$) and their
distance ($D_{ee}$). 
A clear signal of photon conversion is visible in the
$D_{ee}(BP)$-$M_{ee}(BP)$ plane (figure \ref{fig:deemee}).
We reject events having $M_{ee}(BP) < 15\ MeV$ and $D_{ee}(BP) < 2.5\ cm$.
The expected $M_{\pi\pi ee}$ spectrum before and after this cut is shown in
figure \ref{fig:mppee1}.
The $\phi$ background events peaked at the $\eta$ invariant mass are
significantly reduced by this cut.
\\
Finally, to remove continuum background from interactions with quadrupoles,
we have defined the quantities $\ltf$ and $\ltb$ as the average polar angle
of forward and backward particles identified as signal.
Events having  $\ltf > 0.85$ and $\ltb < -0.85$ are rejected.
This cut affects neither the signal nor background from $\phi$
decay events.

\section{Fit to the \mathversion{bold}$\ppee$\mathversion{normal}
         invariant mass spectrum and event counting}
\label{sec:fit-invmass}
\begin{figure}
  \begin{center}
    \includegraphics*[width=5cm]{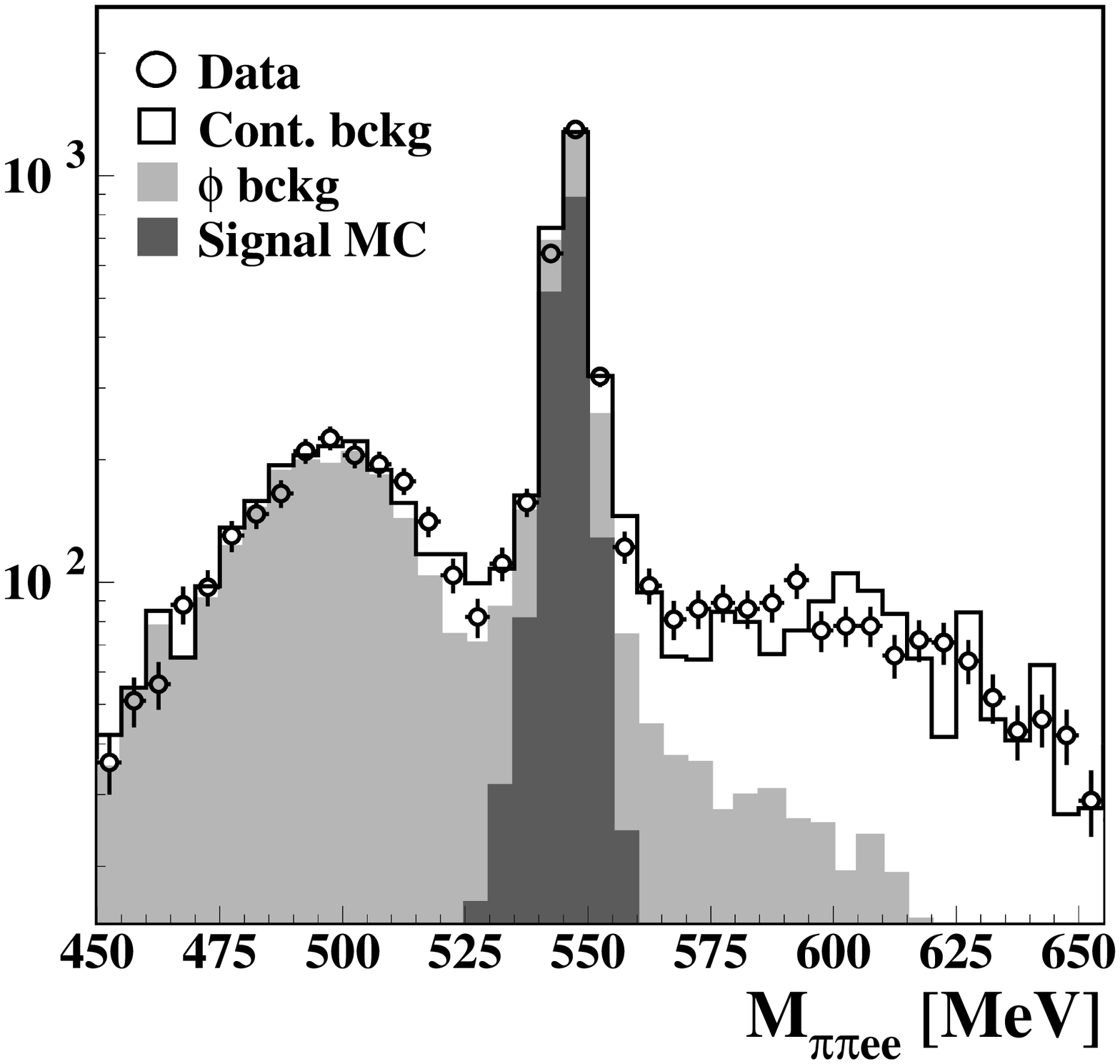}
    \includegraphics*[width=5cm]{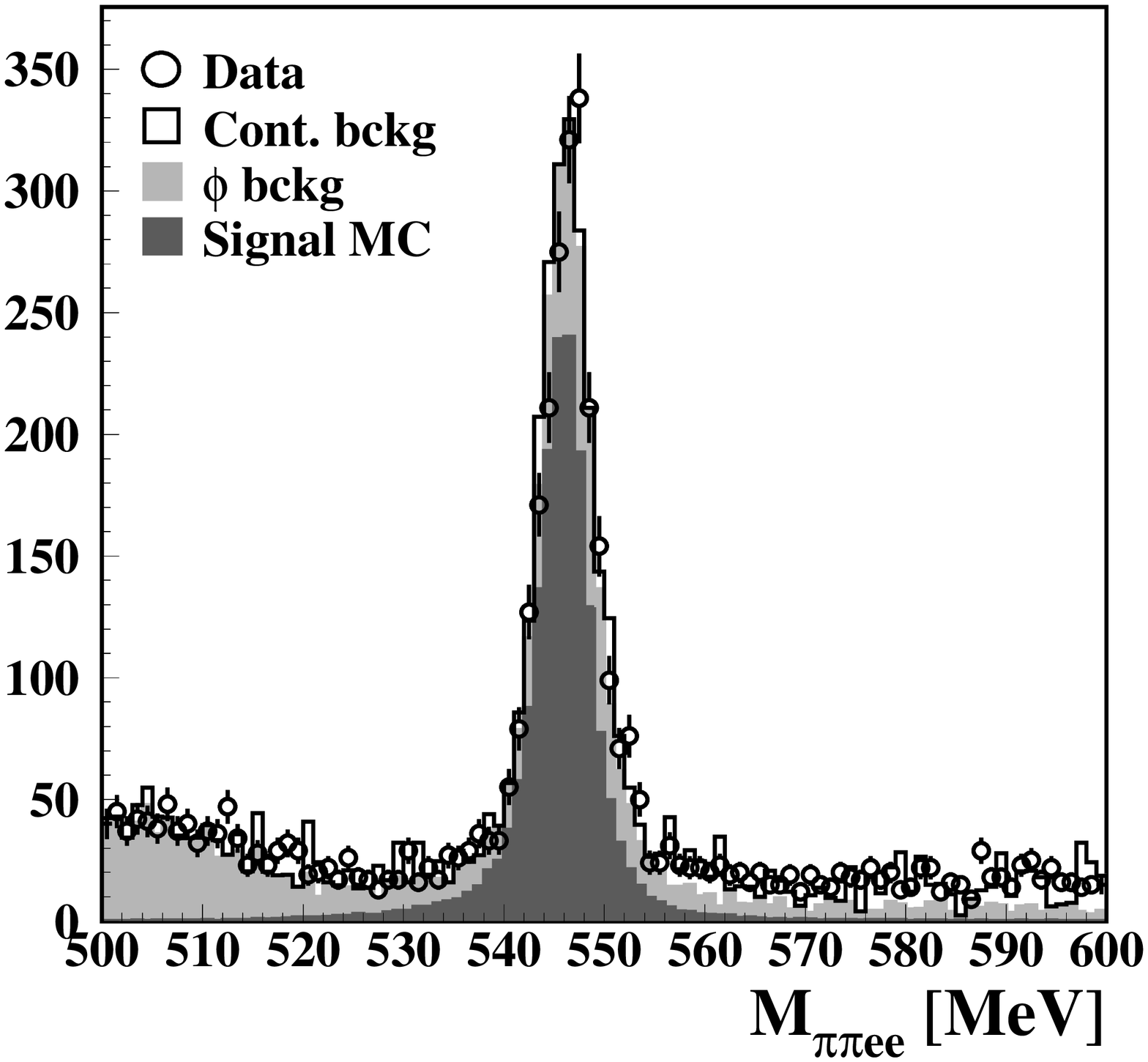}
  \end{center}
  \caption{$\ppee$ invariant mass spectrum on a wide range (left panel) and
           zoomed around the $\eta$ mass (right panel).
           The background scale factors have been obtained as described in
           section \ref{sec:fit-invmass}. 
           Dots: data.
           The black histogram is the expected distribution, i.e. signal MC
           (dark grey), $\phi$ background (light grey) and continuum
           background (white).}
  \label{fig:fitresult}
\end{figure}

\begin{figure}
  \begin{center}
    \includegraphics*[width=5cm]{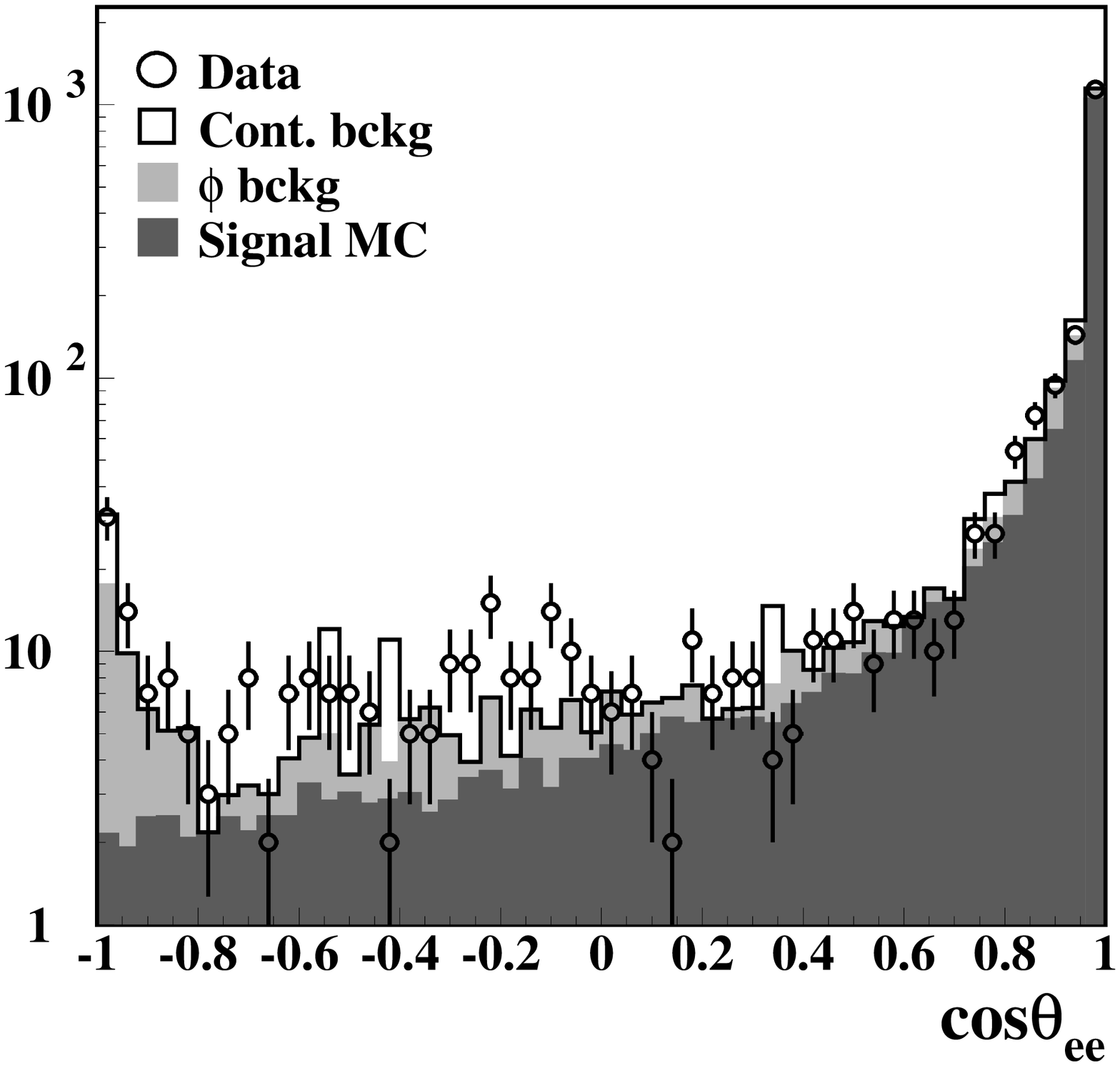}
    \includegraphics*[width=5cm]{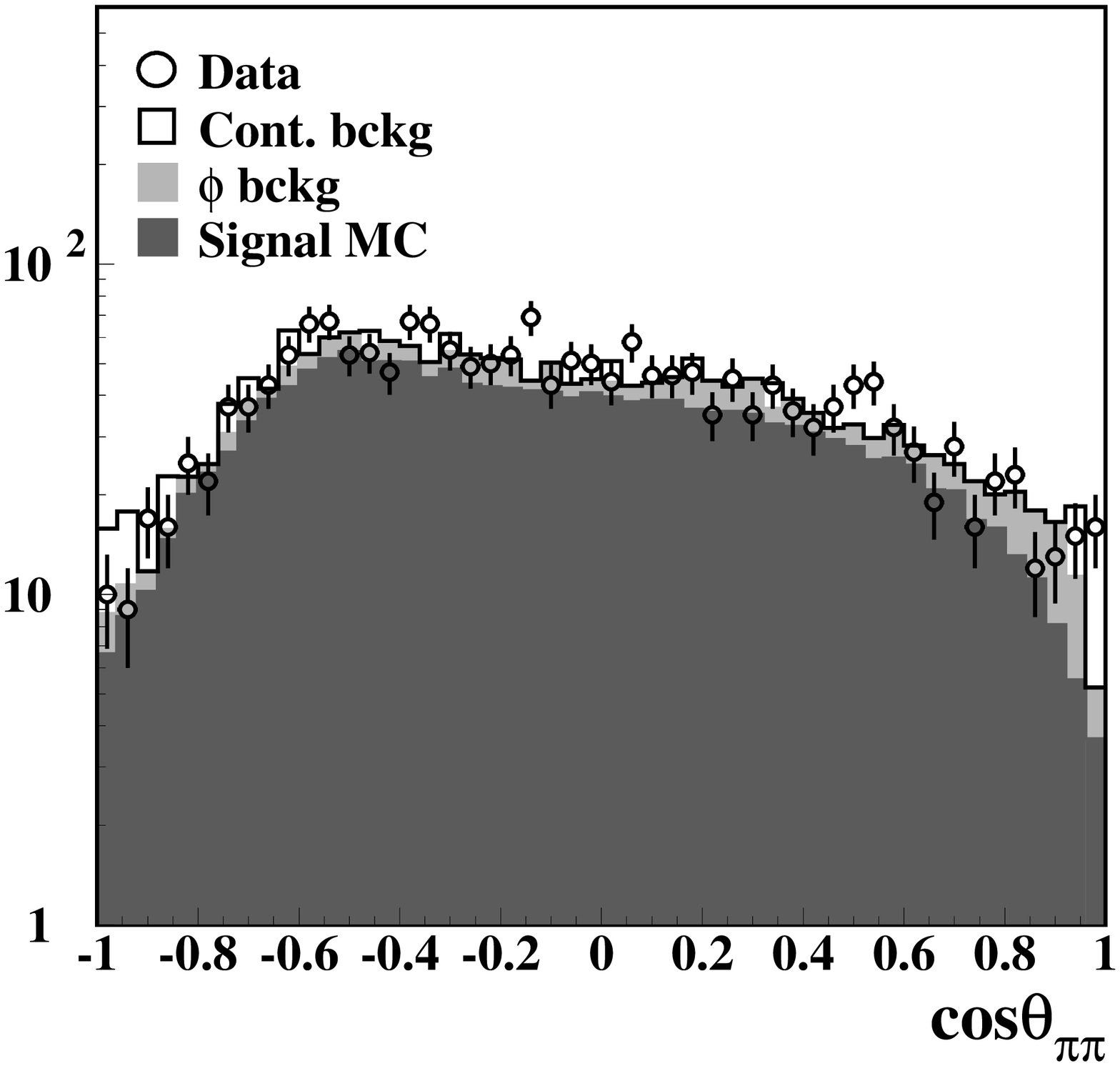}
    \includegraphics*[width=5cm]{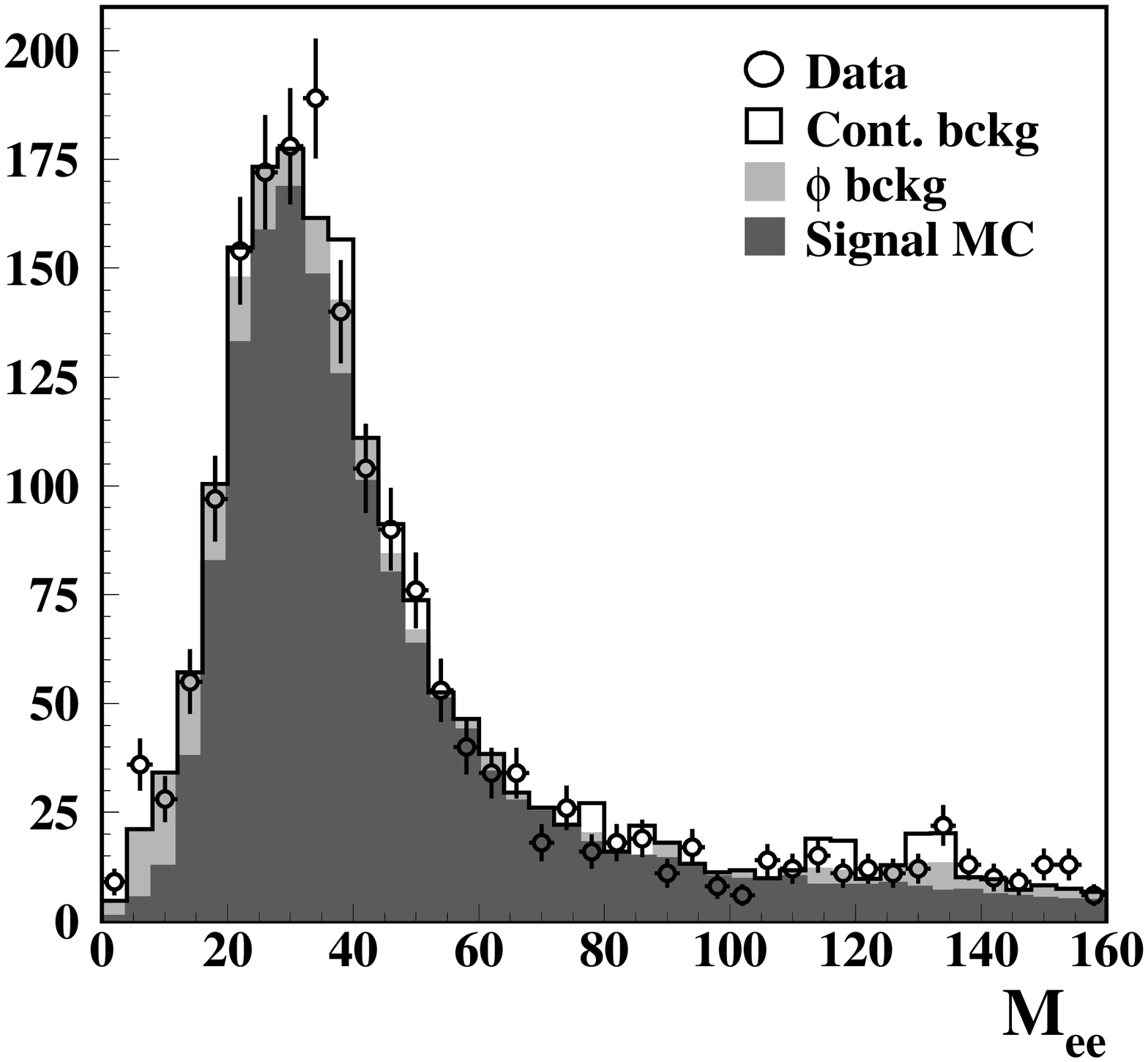}
    \includegraphics*[width=5cm]{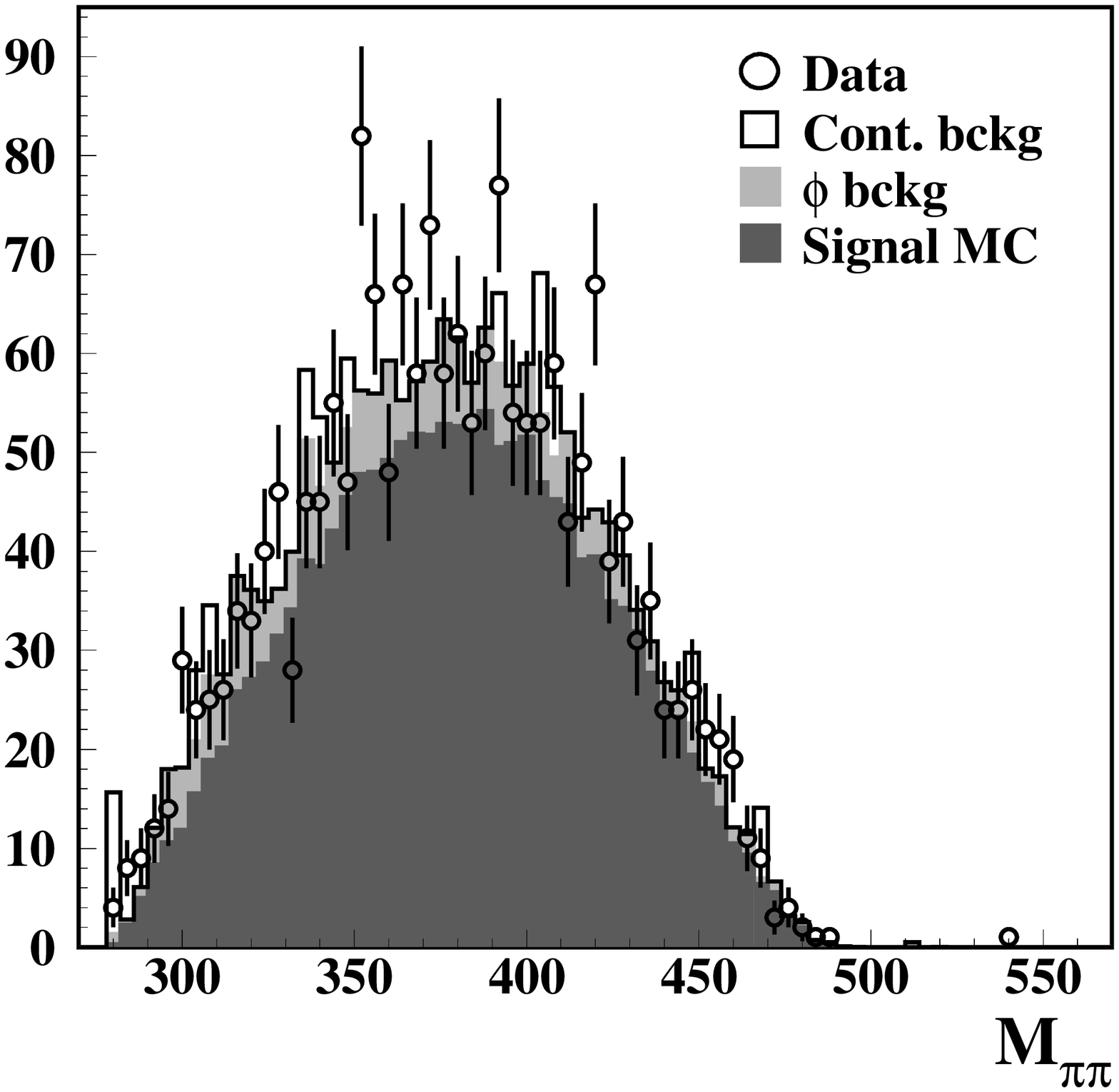}
  \end{center}
  \caption{Spectra of opening angle (top) and invariant mass (bottom) for
           the $\ee$ (left) and $\pp$ (right) pairs for events in the
           signal region.
           The background scale factors have been obtained as described in
           section \ref{sec:fit-invmass}. 
           Dots: data.
           The black histogram is the expected distribution, i.e. signal MC
           (dark grey), $\phi$ background (light grey) and continuum
           background (white).}
   \label{fig:fitcomparison}
\end{figure}

In order to evaluate the background contribution, we perform a fit to the
data distribution of the $\ppee$ invariant mass after the cuts on the
momenta.
The fit is done on sidebands in order not to introduce correlations
between signal and background.
The ranges used are: $[450,520]\ MeV \cup [570,650]\ MeV$.
Upper and lower limits (450 and 650 $MeV$) have been chosen in order not to
include in the fit tails from background distributions.
The central range (520 and 570 $MeV$) is wide enough to well contain the
tails of the signal distribution.
Then the fit is performed using the background shapes only.
\\
The most precise background evaluation has been obtained fixing the
off-peak data scale factor with luminosity at $7.14 \pm 0.03$
and fitting the MC background from $\phi$ meson decay.
The output of the fit is $\chi^2/dof = 32.5/30$ ($P(\chi^2) = 0.35$),
with a scale factor of $0.528 \pm 0.009$, which is in good agreement with
expectation from luminosity. 
The other possible approaches (fixing both the background scale factors
with luminosity, leaving both free in the fit or fixing the
$\phi$ decay and leaving free the continuum background shape) have been
used for the evaluation of systematic uncertainties.
The result of the fit is shown in figure \ref{fig:fitresult} both in a wide
$M_{\pi\pi ee}$ window and around the signal region.
\\
For the signal estimate we limit ourselves to the region $[535,555]\ MeV$
and perform the event counting after background subtraction:
we find 1555 (368) signal (background) events, 
see table \ref{tab:counting}. 
Data-MC comparisons show a very good agreement for all considered
variables.
The most relevant distributions are reported in 
figure \ref{fig:fitcomparison}.  
\begin{table}
  \begin{center}
    \caption{Event counting in the signal region.}
    \label{tab:counting}
    \begin{tabular}{l|r}
      \hline \hline
      Contribution                  & Counting\\
      \hline                        
      Data                          & 1923.0\\
      Signal                        & 1555.0\\
      Total background              &  368.0\\
      $\phi$ background             &  275.2\\
      Continuum background          &   92.8\\
      \hline \hline
    \end{tabular}
  \end{center}
\end{table}

\section{Measurement of the 
         BR(\mathversion{bold}$\etappeeg$\mathversion{normal})}
\label{sec:br-calculation}
The branching ratio has been evaluated according to the formula:
\begin{equation}
  BR(\etappeeg) = \frac{N_{\etappeeg}}{N_{\eta\gamma}} \cdot
                 \frac{1}{\epsilon_{\etappeeg}}
   \label{eq:masterformula}
\end{equation}
where $N_{\etappeeg}$ is the number of signal events and 
$\epsilon_{\etappeeg}$ is the efficiency taken from MC. 
The number of $\phietag$ events, $N_{\eta\gamma}$, has been obtained
using the formula $N_{\eta\gamma} = \mathcal{L} \cdot \sigma_{\phietag}$, 
where $\mathcal{L}$ is the integrated luminosity and the cross section
$\sigma_{\phietag}$ takes into account the $\phi$ meson line shape.
$\sigma_{\phietag}$ has been evaluated with $\etapOpOpO$ events collected
in the range $1017\ MeV < \sqrt{s} < 1022\ MeV$ \cite{csnote}.
Inserting all the numbers summarized in table \ref{tab:summary},
we obtain the value:
\begin{equation}
  BR(\etappeeg) = (26.8 \pm 0.9_{Stat.}) \times 10^{-5}
\end{equation}
where the error accounts for the event counting and the background
subtraction. 
\begin{table}
  \begin{center}
    \caption{Summary of the numbers used in the master formula
             (\ref{eq:masterformula}) for the branching ratio evaluation.}
    \label{tab:summary}
    \begin{tabular}{c|c}
      \hline \hline
      BR inputs                  & Values\\
      \hline                     
      Number of events           & $1555 \pm 52$ \\
      Efficiency                 & $0.0803 \pm 0.0004$ \\
      Luminosity                 & $(1733 \pm 10)\ pb^{-1}$ \\
      $\eephietag$ cross section & $(41.7 \pm 0.6)\ nb$ \\
      \hline \hline
    \end{tabular}
  \end{center}
\end{table}
\\
The systematic uncertainties have been evaluated in the following way:
\begin{itemize}
  \item{fixing with the luminosity or leaving free in the fit the
        background scale factors;}
  \item{varying the sideband upper and lower limits in the ranges 
        $[400,535]\ MeV$ and $[555,700]\ MeV$ respectively;}
  \item{varying the binning of the histogram used for the fit from 1
        MeV/bin to 5 MeV/bin;}
  \item{repeating the whole analysis chain after moving selection criteria 
        on $s2p$, $s4p$, $D_{ee}(BP)$, $M_{ee}(BP)$, $\ltf$, $\ltb$ and
        $M_{\pi\pi ee}$ by $\pm1\sigma$, $\pm2\sigma$'s,
        $\pm3\sigma$'s around the reference value. 
        The BR is then recomputed for all of these variations. 
        The systematic uncertainty has been evaluated as the quadratic sum
        of RMS's obtained for each case.}
\end{itemize}
The uncertainty on $N_{\eta\gamma}$ has been also added to the systematics in
the normalization term.
The results of the systematics evaluation are summarized in table
\ref{tab:systematics}.
The largest contributions are due to the normalization and to the cut on
$M_{ee}(BP)$.
Taking the total systematic error into account, the measurement of the
branching ratio is: 
\begin{equation}
   BR(\etappeeg) = (26.8 \pm 0.9_{Stat.}  \pm 0.7_{Syst.}) \times 10^{-5}\quad.
\end{equation}
\begin{table}
  \begin{center}
    \caption{Summary of the systematic uncertainties on the branching ratio.}
    \label{tab:systematics}
    \begin{tabular}{c|c}
      \hline \hline
      Source of uncertainty    & $\sigma$(BR) \\
      \hline
      Free/fixed scale factors & $0.18 \times 10^{-5}$\\
      Sidebands range          & $0.05 \times 10^{-5}$\\
      Binning                  & $0.02 \times 10^{-5}$\\
      Analysis selection       & $0.55 \times 10^{-5}$\\
      Normalization            & $0.42 \times 10^{-5}$\\
      \hline
      Total                    & $0.72 \times 10^{-5}$\\
      \hline \hline
    \end{tabular}
  \end{center}
\end{table}

\section{Decay plane asymmetry evaluation}
\label{sec:asymmetry}
\begin{figure}
  \begin{center}
    \includegraphics*[width=7cm]{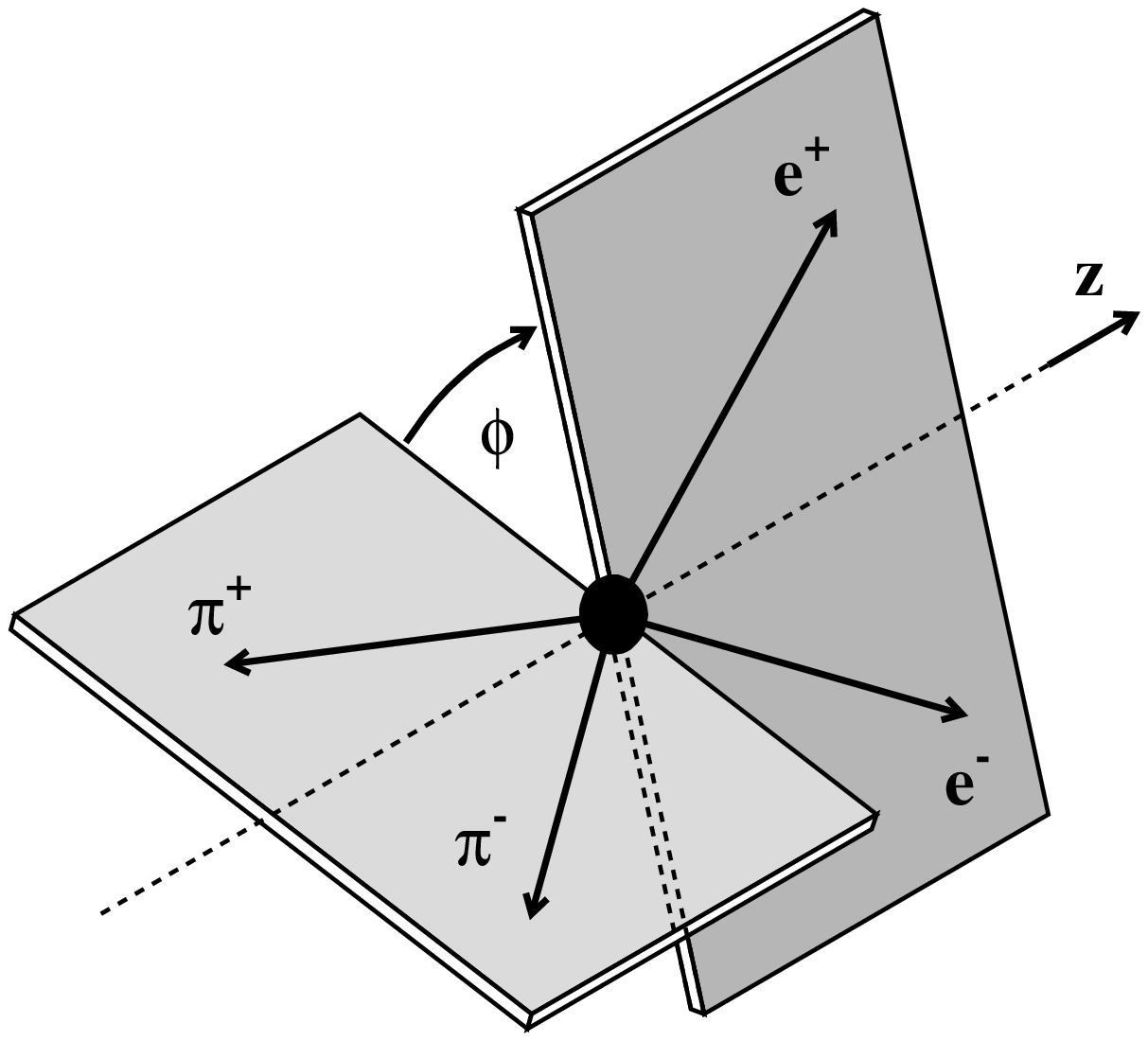}
  \end{center}
  \caption{Definition of the angle $\phi$ between the pion and electron
           decay planes.}
  \label{fig:asym-fig}
\end{figure}

\begin{figure}
  \begin{center}
    \includegraphics*[width=5cm]{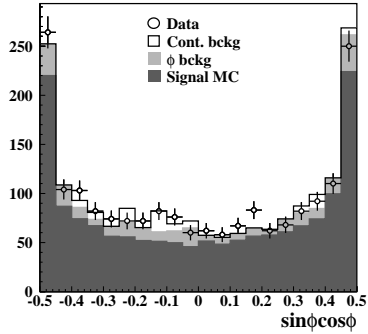}
  \end{center}
  \caption{Distribution of the $\spcp$ variable in the signal region.
           The background scale factors have been obtained as described in
           section \ref{sec:fit-invmass}. 
           Dots: data.
           The black histogram is the expected distribution, i.e. signal MC
           (dark grey), $\phi$ background (light grey) and continuum
           background (white).}
  \label{fig:asym-data}
\end{figure}

The decay plane asymmetry is calculated starting from the momenta of the
four particles and is expressed as function of $\phi$, the angle between
the pion and the electron planes in the $\eta$ rest frame 
(figure \ref{fig:asym-fig}):  
\begin{equation}
  \aphi = \frac{N_{\spcpp}-N_{\spcpm}}{N_{\spcpp}+N_{\spcpm}}\mbox{\quad.}
\end{equation}
The quantity $\sin\phi \cos\phi$ is given by 
$(\hat{n}_{ee}\times\hat{n}_{\pi\pi}) \hat{z} 
 (\hat{n}_{ee}\cdot\hat{n}_{\pi\pi})$,
where the $\hat{n}$'s are the unit normals to the electron and pion planes 
and $\hat{z}$ is the unit vector along the axis defined by the
intersection of the two planes.
The distribution of the $\spcp$ variable in the signal region is shown in
figure \ref{fig:asym-data}. 
We remind that the signal MC has been produced with $\aphi = 0$.
\\
While the analysis efficiency is completely flat in the $\spcp$
distribution, some distortion is introduced by the reconstruction, because
of events with wrong mass assignment.
The correction to this distortion has been evaluated by MC, fitting with a
linear function the ratio between the generated and reconstructed $\spcp$
distributions.
The resulting slope is -0.032 $\pm$ 0.016.
The use of higher polynomials does not improve the fit.
The origin of this slope has been investigated by MC and it is
completely due to the 14\% of signal events with wrong particle
identification.
This because the particle identification algorithm forces the mass
assignment in case of ambiguities without discarding events.
The aim is to preserve the statistics, which completely dominates the
asymmetry measurement.
\\
The asymmetry has been evaluated for the events in the
$535\ MeV<M_{\pi\pi ee}<555\ MeV$ mass region after background subtraction.
After applying the correction, we obtain:
\begin{equation}
  \aphi = ( -0.6 \pm 2.5_{\ Stat.} 
                 \pm 1.8_{\ Syst.}) \times 10^{-2} 
\end{equation}
which is the first measurement of this asymmetry.
\\
As for the branching ratio, the systematic error has been evaluated
repeating the whole analysis chain after varying selection criteria by 
$\pm 1 \sigma$, $\pm 2 \sigma$'s and $\pm 3 \sigma$'s around the reference
value and taking as uncertainty the quadratic sum of the resulting RMS's. 
The uncertainty due to the correction has been evaluated varying its slope
by $\pm 1 \sigma$. 
The largest contribution is due to the cut on $M_{\pi\pi ee}$ while the
contribution of the slope correction is $0.5\times 10^{-2}$.
\\
In order to check the distortion correction applied to $\aphi$, we have
defined a control sample having only events without ambiguities in particle
identification. 
In this case the probability of wrong particle identification is almost
zero and no distortions are observed in the MC $\spcp$ distribution. 
The fraction of this control sample in data and MC events is in good
agreement ($0.62 \pm 0.02$ and $0.64 \pm 0.02$ respectively), showing that
our simulation reproduces the real data well. 
The asymmetry evaluated with the control sample is in good agreement with
our measurement but has a larger statistical error: 
$\aphi = ( -1.2 \pm 3.1_{\ Stat.}) \times 10^{-2}$.

\section{Conclusions}
\label{sec:conclusions}
Using a sample of 1.7 $pb^{-1}$ collected in the $\phi$ meson mass region,
we have obtained a measurement of the $\etappeeg$ branching ratio with 
4\% accuracy, ten times more precise than the previous best measurement
\cite{Akhmetshin:2000bw,Bargholtz:2006gz,Berlowski:2008zz}:
\begin{equation}
   BR(\etappeeg) = (26.8 \pm 0.9_{Stat.}  \pm 0.7_{Syst.}) \times 10^{-5} \quad.
\end{equation}
Radiative events slightly modify momentum distribution of the charged
particles and have been carefully considered in the efficiency evaluation.
As a result, the measured branching ratio is fully radiation inclusive.
\\
Our measurement is about $2\sigma$ smaller than theoretical predictions
\cite{Faessler, Picciotto, Borasoy},
while it is in agreement ($\sim 1\sigma$) with the calculations of the
ratio of the branching fractions $BR(\etappee)/BR(\etappg)$ in references
\cite{Jarlskog, Borasoy} when the recent CLEO measurement of $BR(\etappg)$
\cite{CLEO} is used as normalization. 
\\
The final sample of 1555 signal events allows us to perform the first
measurement of the CP-violating asymmetry $\aphi$, which is consistent with
zero at the 3\% percent precision level:
\begin{equation}
  \aphi = ( -0.6 \pm 2.5_{\ Stat.} \pm 1.8_{\ Syst.}) \times 10^{-2} \quad.
\end{equation}

\section*{Acknowledgements}
\label{acknowledgements}
We would like to thank D.~N.~Gao for the useful discussions.
We thank the DAFNE team for their efforts in maintaining low background
running conditions and their collaboration during all data-taking. 
We want to thank our technical staff: 
G.~F.~Fortugno and F.~Sborzacchi for their dedicated work to ensure an
efficient operation of the KLOE Computing Center; 
M.~Anelli for his continuous support to the gas system and the safety of
the detector; 
A.~Balla, M.~Gatta, G.~Corradi and G.~Papalino for the maintenance of the
electronics;
M.~Santoni, G.~Paoluzzi and R.~Rosellini for the general support to the
detector; 
C.~Piscitelli for his help during major maintenance periods.
This work was supported in part by EURODAPHNE, contract FMRX-CT98-0169; 
by the German Federal Ministry of Education and Research (BMBF) contract
06-KA-957; 
by the German Research Foundation (DFG), 'Emmy Noether Programme', 
contracts DE839/1-4;
and by the EU Integrated Infrastructure Initiative HadronPhysics Project
under contract number RII3-CT-2004-506078.





\begin{thebibliography}{00}





\bibitem{Landsberg}
  L.~G.~Landsberg,
  Phys.\ Rept.\  {\bf 128} (1985) 301.

\bibitem{Jarlskog}
  Jarlskog, C. and Pilkuhn, H.,
  Nucl.\ Phys.\  B {\bf 1} (1967) 264.

\bibitem{Faessler}
  A.~Faessler, C.~Fuchs and M.~I.~Krivoruchenko,
  Phys.\ Rev.\  C {\bf 61} (2000) 035206.

\bibitem{Picciotto}
  C.~Picciotto and S.~Richardson,
  Phys.\ Rev.\  D {\bf 48} (1993) 3395.

\bibitem{Borasoy}
  B.~Borasoy and R.~Nissler,
  Eur.\ Phys.\ J.\  A {\bf 33} (2007) 95.

\bibitem{Akhmetshin:2000bw}
  R.~R.~Akhmetshin {\it et al.} [CMD-2 Collaboration],
  Phys.\ Lett.\ B {\bf 501} (2001) 191.

\bibitem{Bargholtz:2006gz}
  C.~Bargholtz {\it et al.} [CELSIUS-WASA Collaboration],
  Phys.\ Lett.\ B {\bf 644} (2007) 299.

\bibitem{Berlowski:2008zz}
  C.~Berlowski {\it et al.},
  Phys.\ Rev.\ D {\bf 77}, 032004 (2008).


\bibitem{Sehgal:1992wm}
  L.~M.~Sehgal and M.~Wanninger,
  Phys.\ Rev.\  D {\bf 46} (1992) 1035
  [Erratum-ibid.\  D {\bf 46} (1992) 5209].

\bibitem{Heiliger:1993qt}
  P.~Heiliger and L.~M.~Sehgal,
  Phys.\ Rev.\  D {\bf 48} (1993) 4146
  [Erratum-ibid.\  D {\bf 60} (1999) 079902].

\bibitem{Elwood:1995xv}
  J.~K.~Elwood, M.~B.~Wise and M.~J.~Savage,
  Phys.\ Rev.\  D {\bf 52} (1995) 5095
  [Erratum-ibid.\  D {\bf 53} (1996) 2855].

\bibitem{Elwood:1995dj}
  J.~K.~Elwood, M.~B.~Wise, M.~J.~Savage and J.~W.~Walden,
  Phys.\ Rev.\  D {\bf 53} (1996) 4078.

\bibitem{Ecker:2000nj}
  G.~Ecker and H.~Pichl,
  Phys.\ Lett.\  B {\bf 507} (2001) 193.


\bibitem{AlaviHarati:1999ff}
  A.~Alavi-Harati {\it et al.}  [KTeV Collaboration],
  Phys.\ Rev.\ Lett.\  {\bf 84} (2000) 408.

\bibitem{Lai:2000xf}
  A.~Lai {\it et al.}  [NA48 Collaboration],
  Phys.\ Lett.\  B {\bf 496} (2000) 137.


\bibitem{Ambrosino:2004ww}
  F.~Ambrosino {\it et al.}  [KLOE Collaboration],
  Phys.\ Lett.\  B {\bf 606} (2005) 276.

\bibitem{Jarlskog:1995ww}
  C.~Jarlskog and E.~Shabalin,
  Phys.\ Rev.\  D {\bf 52} (1995) 248.

\bibitem{Geng:2002ua}
  C.~Q.~Geng, J.~N.~Ng and T.~H.~Wu,
  Mod.\ Phys.\ Lett.\  A {\bf 17} (2002) 1489.

\bibitem{Gao:2002gq}
  D.~N.~Gao,
  Mod.\ Phys.\ Lett.\  A {\bf 17} (2002) 1583.


\bibitem{DCH}
  M.~Adinolfi {\it et al.}  [KLOE Collaboration],
  Nucl.\ Inst.\ and Meth.\ A 488 (2002) 51.
\bibitem{EMC}
  M.~Adinolfi {\it et al.}  [KLOE Collaboration],
  Nucl.\ Inst.\ and Meth.\ A 482 (2002) 364.
\bibitem{TRG}
  M.~Adinolfi {\it et al.}  [KLOE Collaboration],
  Nucl.\ Inst.\ and Meth.\ A 492 (2002) 134.
\bibitem{NIMOffline}
  F.~Ambrosino {\it et al.}  [KLOE Collaboration],
  Nucl.\ Inst.\ and Meth.\ A 534 (2004) 403.

\bibitem{Barberio:1993qi}
  E.~Barberio and Z.~Was,
  Comput.\ Phys.\ Commun.\  {\bf 79} (1994) 291.

\bibitem{Golonka:2005pn}
  P.~Golonka and Z.~Was,
  Eur.\ Phys.\ J.\  C {\bf 45} (2006) 97.

\bibitem{csnote}
  S.~Giovannella and S.~Miscetti,
  KLOE note 212 (2006), \\
  http://www.lnf.infn.it/kloe/pub/knote/kn212.ps .

\bibitem{CLEO}
  A.~Lopez {\it et al.}  [CLEO Collaboration],
  Phys.\ Rev.\ Lett.\  {\bf 99} (2007) 122001.


\end{thebibliography}
\end{document}